\documentstyle[preprint,aps]{revtex}
\begin{document} 
\draft 
\title{ 
Frustrated quantum
Heisenberg ferrimagnetic chains} 
 
\author{N. B. Ivanov } 
\address{Institute  for Solid  State 
Physics, Bulgarian Academy of Sciences,\\ 
Tzarigradsko 
chaussee-72, 1784 Sofia, Bulgaria} 
\author{J. Richter} 
\address{Institut f\"ur Theoretische Physik, 
Universit\"at Magdeburg,\\ 
 P.O.Box 4120, D-39016 Magdeburg, Germany} 
\author{U. Schollw\"ock} 
\address{Sektion Physik, Ludwig-Maximilians-Universit\"at M\"unchen,\\ 
Theresienstrasse 37/III, 80333 Munich, Germany\\} 
 
\maketitle 
\begin{abstract} 
 
We study the ground-state properties of  weakly frustrated 
Heisenberg ferrimagnetic chains with nearest 
and next-nearest neighbor antiferromagnetic 
exchange interactions and two types of alternating 
sublattice spins $S_1>S_2$, using $1/S$ spin-wave 
expansions, density-matrix renormalization group, and 
exact-diagonalization techniques. 
It is argued that the zero-point spin fluctuations 
completely destroy the 
classical commensurate-incommensurate continuous 
transition. Instead, the  long-range ferrimagnetic state
disappears through a discontinuous
transition to a singlet state at a larger  value
of the frustration parameter.
In the ferrimagnetic phase we find
a disorder point marking the onset of incommensurate
real-space short-range spin-spin correlations. 
\end{abstract} 
\vspace{0.3in} 
\section{Introduction}

Over the last years a great deal of interest 
has been concentrated on 
the physics of quantum Heisenberg spin chains with competing 
exchange couplings both for half-integer and integer site
spins (see, e.g., Refs.\onlinecite{bursill,scholl,wite} and the 
references therein). In general, frustration reduces 
the antiferromagnetic correlations and in some cases may 
produce various exotic quantum ground states such as the 
dimerized state, the spin-nematic state, 
or some sort of spin-liquid 
states. At the same time, little is known about 
the role of frustration in the so-called mixed spin chains 
constructed from two or more kinds of site spins regularly 
distributed on the lattice. Recently,  a number of studies 
concerning the physics of unfrustrated  mixed 
Heisenberg chains with two kinds of spins 
have been published\cite{pati,tian,fukui,xian}. Depending on the
periodic array, the mixed models can form massive or 
massless ground states and display a rich variety of
new features\cite{fukui,xian}. Meanwhile, 
various mixed one-dimensional quantum spin 
systems have already been synthesized 
in the last decade\cite{kahn}. 
 
In this report we address  one-dimensional 
spin chains  containing two different 
alternating site spins $S_1>S_2$ 
per unitary cell and interacting via 
competing antiferromagnetic neighbor and next-neighbor 
couplings, $J_1>0$ and $J_2>0$, Fig.1. 
The Hamiltonian of the system reads: 
\begin{equation} 
H= J_1\sum_{n,\delta}{\bf S_1}_{n}{\bf S_2}_{n+\delta}+ 
J_2\sum_{n}\left( {\bf S_1}_{n}{\bf S_1}_{n+1}+ 
{\bf S_2}_{n}{\bf S_2}_{n+1}\right), 
\label{ham} 
\end{equation} 
where the  integers $n$ number 
$N$ unitary cells and the vector $\delta=\pm 1/2$ 
connects the $S_1$ spins with the 
nearest-neighbor $S_2$ spins. The size of 
the elementary cell is unity. 
In what follows we frequently use the notations 
 $w \equiv S_1/S_2$, $S \equiv S_2$, $J \equiv J_2/J_1$, 
$J_1\equiv 1$. 
Theoretical models of quantum ferrimagnetic systems with 
competing interactions have already been discussed in 
the literature\cite{vega}. However, these models consider 
complicated multiple spin-spin interactions which are 
far from the real experiment. On the other hand, 
little is known about the 
simple frustrated Heisenberg models
containing only two-spin couplings.
 
In the classical limit, the ground state of Eq.(\ref{ham}) 
can be described by the ansatz: 
\begin{equation} 
{\bf S_i}_{n}=S_i\left[\hat{\bf u } \cos(Qn)+ 
\hat{\bf v }\sin(Qn)\right],\hspace{0.5cm} i=1, 2, 
\end{equation} 
where $\hat{\bf u }\perp \hat{\bf v }$ are unit 
vectors in the spin space. The classical 
commensurate ferrimagnetic state with a pitch  angle 
between the nearest spins $\theta \equiv Q/2=\pi$ 
is stable up to  the phase transition 
point 
\begin{equation} 
J_C=\frac{ w}{2(w^2+1)}. 
\label{tp} 
\end{equation} 
For larger values of frustration, $J>J_C$, 
the stable state is a spiral state with an ordering wave vector given by 
\begin{equation} 
\cos\frac{Q}{2}=-\frac{w}{2J (w^2+1)}. 
\label{Q} 
\end{equation} 
In the limit $J = \infty$, $\theta=\pi/2$ and 
the system is composed of two decoupled 
antiferromagnetic chains with site spins $S_1$ and $S_2$. 
The  long-range N\'eel
order is excluded in this limit, but 
one might expect the classical 
result to give a guide to possible finite-range order.
In the extreme antiferromagnetic case, $w=1$, Eq. (\ref{Q}) 
reduces to the well-known result 
for homogenous frustated antiferromagnetic 
chains, and $J_C$ reaches its upper bound $1/4$. 
Recent studies 
of the $J =0$ system  confirm the expectation that 
the classical ferrimagnetic state survives the quantum spin 
fluctuations\cite{pati}. Note that the ferrimagnetic
state is characterized by both longitudinal ferromagnetic
and antiferromagnetic long-range orders\cite{tian}.
Thus, a special feature of
the discussed model, Eq. (\ref{ham}), is the existence of 
an order-disorder quantum phase transition. The character 
of the latter transition should crucially depend on 
the values of  quantum  site spins $S_1$ and $S_2$. 
 
\section{ Spin-wave analysis}
 
In this report we consider relatively small values 
of the frustration parameter $J$. 
In the unfrustrated case, the linear spin-wave 
theory (LSWT) 
has already been successfully 
applied in a series of recent works\cite{pati}. 
The second-order $1/S$ series for a number of quantities 
were shown\cite{ivanov} to reproduce with a high precision the 
density matrix renormalization group (DMRG) results. 
Thus, we can expect that at least in the weakly frustrated 
region the spin-wave theory (SWT) gives a realistic 
picture of the low-energy physics of the system. 
However, its application to other frustrated spin 
systems reveals that in many cases 
the linear theory gives better qualitative results\cite{chandra}. 
In this respect, we observe similar behavior in the present 
system, 
so that  the use of spin-wave expansions  requires some care. 
 
In the ferrimagnetic region 
LSWT predicts the existence of two types 
of elementary excitations: 
\begin{equation} 
E_k^{(a,b)}= 
2S\left(\frac{a_k+b_k}{2}\epsilon_k\pm 
\frac{(a_k-b_k)}{2}\right), 
\label{disp} 
\end{equation} 
where $a_k=1-w(1-\cos k)$, $b_k=w-J (1-\cos k)$, 
$\epsilon_k=\sqrt{1-\eta_k^2}$, and $\eta_k=2\sqrt{w} \cos(k/2)/ 
(a_k+b_k)$.
The $E_k^{(a)}$ excitations are gapless, $E_k^{(a)}\sim 
k^2$ for small $k$. They describe ferromagnetic  magnons 
in the sector with a total 
spin $(S_1-S_2)N-1$. 
The $E_k^{(b)}$ excitations are gapful, $E_k^{(b)}= 
\Delta+O(k^2)$, and belong to  the sector $(S_1-S_2)N+1$ 
(antiferromagnetic magnons). 
The existence of gapless excitations at $k=0$ reflects the 
continuous symmetry of the Hamiltonian under a global rotation 
of spins. On the other hand, for $w>1$ the antiferromagnetic 
excitations acquire a gap due to the violated sublattice symmetry. 
In the weakly frustrated region the  structure of low-lying 
energy levels can also 
be predicted from the Lieb-Mattis theorem\cite{lieb}: 
For every finite $N$, the energy levels
order according to the rule:
\begin{equation} 
E(S_t+1) > E(S_t), \hspace{0.5cm}, S_t \geq S_g, 
\end{equation} 
\begin{equation} 
E(S_t) > E(S_g), \hspace{0.5cm}, S_t < S_g, 
\end{equation} 
where $S_t$ is the total spin of the state and 
$S_g=(S_1-S_2)N$ is the spin of the ground state. 
Strictly speaking, the Lieb-Mattis theorem does not 
work in the frustrated system, but due to the continuity 
principle we can expect 
that the order of states survives up to some finite 
$J$\cite{richter}. The above picture is also confirmed 
in our exact-diagonalization studies of small clusters. 
On the other hand, the strong frustration might 
cause a substantial disarrangement of levels, as 
is shown numerically below. 
 
Exactly at the classical transition 
point $J = J_C$,  the  quadratic term in the 
small $k$ expansion of $E_k^{(a)}$ vanishes, so that 
the spectrum reads: 
$E_k^{(a)}=ck^4+O(k^6),\hspace{0.5cm} c>0$, $J =J_C$.
For $J > J_C$, 
the classical ground state is the spiral state described 
above, Eq.(\ref{Q}). Now  LSWT predicts  zero modes
 at $k=0$ and at $k=\pm (2\pi -Q)$.
However, the classically broken
SO(3) symmetry in the spiral state is generally expected
to be restored by quantum fluctuations in one-dimensional
systems\cite{azaria}, so that a crucial role of
boson-boson interactions
should be expected near and beyond the classical transition 
point.
 
It is instructive to trace back the change in the 
gap function $\Delta(J)$. 
The first two terms in the $1/S$ series are:
\begin{equation} 
    \Delta=(w-1)\left( 2S-\frac{2g_1}{\sqrt{w}}\right) 
    +O\left(\frac{1}{S}\right), \hspace{0.5cm} 
    g_1=-\frac{1}{2N}\sum_k\frac{\eta_k}{\epsilon_k}\cos 
    \frac{k}{2}. 
\end{equation} 
In Fig. 2 we show the dispersion functions, Eq. (\ref{disp}), 
in the case $(S_1,S_2)= 
(1,1/2)$  for different frustration parameters $J$. 
Near the transition 
point the ferromagnetic mode 
is strongly  flatted, whereas 
the changes in $E_k^{(b)}$ are modest. In addition, 
Fig. 3 displays 
a smooth increase of the gap $\Delta$ 
up to the classical transition point 
$J_C$, which means that the antiferromagnetic 
excitations do not play any important role in the mechanism 
of the transition.
 
In Figs. 4 and  5  second-order SWT 
results for the ground state energy 
and the sublattice magnetization are compared to those of the DMRG 
and the exact diagonalization (ED) 
numerical methods. SWT values for the energy 
are close to the numerical ones in a large region 
of the ferrimagnetic phase. As to the magnetization, SWT is 
effective only in a small vicinity of the unfrustrated model. 
Such a collapse of spin-wave series has already been indicated 
in other frustrated spin models\cite{igarashi}.
We are faced with an example where the linear theory 
gives a better qualitative 
description. 
 
Let us now address the spin-spin correlations. 
The latter will be analysed quantitatively 
in the framework of the DMRG method below. Here the 
purpose is to demonstrate that LSWT qualitatively captures 
the asymptotic behavior of the finite-range 
spin-spin correlations. 
As an example, we consider the
spin-spin correlator 
$K_{12}^{ \perp }(r) = \langle 
{\bf S_1^+}_{n}{\bf S_2^-}_{n+r}+ 
{\bf S_1^-}_{n}{\bf S_2^+}_{n+r} 
\rangle /2$. In a linear 
spin-wave approximation we have: 
\begin{equation} 
K_{12}^{\perp}(r)=-\frac{2S\sqrt{w}}{N} 
\sum_k\frac{\eta_k}{\epsilon_k}\cos(kr). 
\label{corr} 
\end{equation} 
The asymptotic behavior of $K_{12}^{\perp}(r)$ for
long distances $r$ depends on the analytic 
structure of the dominator in 
Eq. (\ref{corr}). It is easy to check that in the region 
$ 0 < J < J_D$, 
\begin{equation} 
J_D = \frac{1}{2(w+1)},
\label{dop} 
\end{equation} 
 the poles in 
Eq. (\ref{corr}) are pure imaginary. Using an appropriate 
standard integral, on obtains the following  two-dimensional 
(2D) Ornstein-Zernike type asyptotic expression: 
\begin{equation}
K_{12}^{\perp}(r) \sim \frac{e^{-\frac{r}{\xi}}}{\sqrt{r}}, 
\hspace{0.5cm} r \rightarrow \infty, \hspace{0.5cm} 
J < J_D.
\label{k12} 
\end{equation} 
This can be generically expected for a one-dimensional 
quantum problem from field-theoretical arguments. 
$\xi$ is the correlation length of the short-range transverse 
spin fluctuations. In the limit $J = 0$, LSWT predicts
$\xi = 1/ \ln w$, as it was also obtained by Pati et al.\cite{pati}. 
The LSWT function $\xi(J )$, $0\leq J\leq J_D$, 
compared to the DMRG results, is shown in Fig. 6. 
A well-pronounced effect of fructration 
is the reduction of the correlation length. Qualitatively, 
LSWT reproduces the latter tendency. 
The special point $J_D$, known as a disorder point 
of the first kind in classical thermodynamics, 
marks the onset of incommensurate 
finite-range spin-spin correlations in the chain 
(see, e.g., 
Ref.\onlinecite{scholl} and references 
therein). In the region $J > J_D$ the poles in 
Eq. (\ref{corr}) move away from the imaginary axis. 
Due to the emergence of a real part, 
incommensurate real-space spin-spin correlations appear. 
Later on, using also precise DMRG data, 
we shall discuss in more detail 
the physics of the disorder point $J_D$. 
 
\section{ DMRG and exact-diagonalization analysis}
 
To obtain quantitative results, we have used  DMRG 
and  Lanczos exact-diagonalization methods. DMRG is 
particularly adapted to the problem, for it allows 
treating large enough frustrated systems. 
DMRG is also 
free from the negative sign problems characteristic 
of Monte Carlo methods when applied to frustrated systems. We have studied 
three types of open spin chains, $(S_1,S_2)=(3/2,1)$, $(1,1/2)$, 
and $(3/2,1/2)$, containing up to $N=20$ unit cells, many times the 
observed correlation lengths. 
The DMRG results for the ground-state energy 
and sublattice magnetization are presented in Table 1.
The sublattice magnetization, measuring the long-range
antiferromagnetic order in the ferrimagnetic state, displays
only a slight monotonic decrease and remains finite up to
the transition point $J_T$ discussed below.
 
An analysis of the correlations calculated 
by  DMRG for large values of frustration is 
difficult due to incommensurability. As to 
the small frustration region, the behavior 
of the transverse spin-spin correlations 
can clearly be extracted 
for distances  $r$ up to $20$ ( the limit is due to the very small 
correlation length $\xi$, leading to a very fast decay 
to numerical noise). Fits clearly reveal that
the planar correlations obey the $2D$ Ornstein-Zernike 
form, Eq. (\ref{k12}), in accord to the qualitative 
LSWT analysis. The fit can be done over many orders of 
magnitude. A fit to a purely exponential law is much worse. 
The correlation lengths given in Table 2 
are extracted from these fits. As an example, 
in Fig. 6 we show the correlation length $\xi$ vs. $J$
in the case $(1,1/2)$. 
We recognize a typical behavior for the function $\xi(J)$, 
showing the existence of a 
disorder point at $J \approx 0.177$.
Recently, it has been argued by one of the authors\cite{scholl}, 
that in quantum spin chains at $T=0$ the presence of a phase 
transition from a commensurate to an incommensurate phase 
in the classical limit should be generically reflected 
in the presence of so-called disorder points, a concept 
of classical statistical mechanics. Indeed, such disorder 
points have already been identified in a variety of cases, 
such as frustrated spin chains, as well as in the 
AKLT model\cite{scholl}. We have clearly found such quantum 
remnants of the classical phase transition also in the 
present model, Table 3. The point $J_D$ 
is characterized by the onset of incommensurate real-space 
correlations as well as by an infinite derivative 
of the correlation length $d\xi /dJ$ at 
$J =J_D-0$. We were not able to extract conclusively 
at the disorder point the crossover from two-dimensional 
to one-dimensional Ornstein-Zernike correlation functions, 
due to the extreme shortness of the correlation length. 
Generically, it can be expected that the disorder point shifts 
to smaller values of frustration for larger $w$, as it
has been borne out by our numerical results as well as by 
the spin-wave analysis, Eq. (\ref{dop}). 
 
Not as satisfying data could be obtained  
for the $\langle S^zS^z\rangle$-correlations: As 
we are working in a fixed $S^z_{total}\neq 0$ environment, 
we have to subtract one-site expectation values $\langle S^z \rangle$.
These values introduce substantial errors, 
given that the actual correlation is smaller by orders of 
magnitude than the subtracted value; in addition, 
the longitudinal correlation length is shorter than the 
transverse one. 
 
Finally, let us consider the structure of the low-energy 
levels around the classical transition as obtained from 
the DMRG and ED simulations. 
Both methods 
seem to be  effective 
tools revealing a complicated picture of low-energy level crossings.
At the transition point $J_T>J_C$ (see Table 3) a first-order
transition (level crossing) takes place from the ferrimagnetic
state with $S_g=(S_1-S_2)N$ to a singlet $S_g=0$ state.
This is accompanied by a basic rearrangement of the spin correlations.
Beyond the transition point $J_T$, the ED numerical analysis
indicates additional level crossings. To be specific,
we illustrate that for a periodic system
with $N=6$ cells and $(S_1,S_2)=(1,1/2)$:
The discontinuous transition from the ferrimagnetic
ground state with total spin
$S_g=(S_1-S_2)N=3$ to the ground state
with $S_g=0$ appears at $J=0.229$. This singlet ground
state is stable up to $J=0.260$. In the next interval
$(0.260,0.303)$ the lowest energy level is a triplet, $S_g=1$.
Beyond the point $J\approx 0.303$ the ground state has
$S_g=0$, but there are several crossings between singlets
of different symmetries at $J\approx 0.417$, $0.735$,
and $0.759$. These may however be due to incommensurability
effects of short chains. DMRG shows that at least from
$J=0.25$ onwards the ground state has $S_g=0$; for values
between the transition and $J=0.24$, numerically a $S_g=1$
ground state cannot be clearly excluded due to a very
small spacing between the levels just above the transition
and at the same time a large DMRG error close to the transition.
Definitely, no higher $S_g$ is observed for the ground state.
Together with the ED result, we believe however that the
ground state is indeed $S_g=0$.
 
\section{ Discussion}
 
In this paper the emphasis was on relatively 
small values of the frustration parameter $J$, when the 
classical ferrimagnetic state survives the zero-point 
spin fluctuations. As a whole, frustration produces 
a strong reduction of the finite-range spin-spin correlation 
length. Our key results are: 
(i) The quantum spin fluctuations 
destroy the classical continuous transition and generate a 
discontinuous transition from the long-range ordered
ferrimagnetic
$S_g=(S_1-S_2)N$ state to a singlet state, $S_g=0$.
The transition point $J_T$ lies beyond the classical
transition point $J_C$.
(ii) The discontinuous transition point $J_T$ is preceded by
a disorder point $J_D<J_T$, marking
the onset of incommensurate 
real-space short-range correlations. For smaller 
values of frustration, $J<J_D$, the transverse 
short-range spin-spin correlations show 
typical $2D$ Ornstein-Zernike behavior. 
 
Even in the small $J$ region, a number
of open issues can be indicated. One of the
important questions concerns
the nature of the $S_g=0$ ground states
established  beyond the ferrimagnetic phase.
As mentioned above, the classically broken
SO(3) symmetry in the spiral state is generally expected
to be restored by quantum fluctuations in the one-dimensional
systems. Thus, we are enforced to look for
possible magnetically disordered states. A valuable
information can be obtained from the Lieb-Schultz-Mattis
theorem\cite{lieb2} adapted to
mixed chain spin models\cite{fukui}.
In our case, the theorem is applicable to the systems
$(S_1,S_2)=(1,1/2)$ and $(3/2,1)$, and
says that  the model (\ref{ham}) either has gapless
excitations or else has degenerate ground states.
Thus, we should  look for phases with presumably
some discrete symmetry broken. An analysis of
possible quantum spin phases has already been
implemented in the context of the model of
frustrated ferromagnetic
chains ($J_1<0$, $J_2>0$)\cite{chubukov,krivnov}.
In particular, two possible nematiclike
phases with additional symmetry breaking of
reflections about a bond or about a site have been
suggested\cite{chubukov}. It was argued above that
the antiferromagnetic excitations in our model do
not play any important role  near  $J_T$
due to the large energy gap $\Delta$ near the transition.
Therefore, the  instability of  the ferrimagnetic
state is expected to capture the main features
of the ferromagnetic instability appearing
in  the frustrated ferromagnetic chain. However,
in contrast to the cascade of transitions predicted
for the ferromagnetic chain\cite{krivnov}, our numerical
results point towards a direct transition from the
ferrimagnetic $S_g=(S_1-S_2)N$ state to a state
with $S_g=0$. Such a difference in the behavior of
both models at $J_T$ may be
attributed to the low-energy multimagnon
bound states appearing in the ferromagnetic chain.
In addition,  the variety of possible nonmagnetic
quantum phases beyond $J_T$ may be considerably enlarged
due to the existence of two kinds of site
spins, $S_1$ and $S_2$, in the mixed chain model.
The complicated picture of low-energy level
crossings described above points
towards a rich
phase diagram of the discussed ferrimagnetic model, as well.
Studies in this direction are in progress at the moment.
 
\acknowledgements

One of the authors (N.B.I.) thanks 
the staff of the Institut f\"{u}r Theoretische Physik, Magdeburg, 
for hospitality. The stay in the Universit\"{a}t Magdeburg 
was supported by the Deutsche Forschungsgemeinschaft (436 BUL
17/9/97). J.R. thanks the Deutsche Forschungsgemeinschaft
for support (project Ri 615/1-2).

 
\pagebreak

\begin{figure} 
\caption{  The mixed-chain Heisenberg 
spin model studied in the paper. $J_1$, $J_2>0$.} 
\label{fig1} 
\end{figure} 
 
\begin{figure} 
\caption{  The dispersion functions $E^{(a,b)}_k$
in a first-order spin-wave approximation for different 
values of the frustration parameter $J$.} 
\label{fig2} 
\end{figure} 
 
\begin{figure} 
\caption{  The antiferromagnetic gap $\Delta $ 
vs. $J$ obtained in a first-order 
spin-wave approximation (solid curves) and by the Lanczos 
exact-diagonalization method.} 
\label{fig3} 
\end{figure} 
 
\begin{figure} 
\caption{ 
The ground-state energy per cell $E_0$ 
of the frustrated $(1,1/2)$ ferrimagnetic chain 
obtained by the second-order SWT, DMRG, and 
the exact-diagonalization methods.} 
\label{fig4} 
\end{figure} 
 
\begin{figure} 
\caption{  On-site magnetization of the first sublattice $m_1$ 
in the  frustrated $(1,1/2)$ ferrimagnetic chain 
obtained by LSWT, DMRG, and 
the exact-diagonalization methods. The second-order spin-wave 
results are not conclusive, as explained in the text.} 
\label{fig5} 
\end{figure}

\begin{figure} 
\caption{  The transverse spin-spin correlation length in 
the frustrated $(1,1/2)$ system vs. $J$ 
obtained from the LSWT and the DMRG method. 
$J_D\approx 0.177$ is the disorder point marking the
onset of incommensurate real-space  finite-range 
spin-spin correlations.} 
\label{fig6} 
\end{figure} 
\newpage
\begin{table} 
\caption{ DMRG results for the ground-state energy per cell $E_0$, 
and the on-site sublattice magnetization $m_1$ 
of the first sublattice in the systems $(S_1,S_2)=(3/2,1)$, 
$(1,1/2)$, and $(3/2,1/2)$. } 
\label{table1} 
\begin{tabular}{ccccccc} 
&\multicolumn{2}{c}{$(\frac{3}{2},1)$} 
&\multicolumn{2}{c}{$(1,\frac{1}{2})$} 
&\multicolumn{2}{c}{$(\frac{3}{2},\frac{1}{2})$}\\ 
 
$J$ &$E_0$      &$m_1$      & $E_0$        &$m_1$        &$E_0$ 
&$m_1$ \\ 
\tableline 
0.00   &-1.93096  &1.14428  &-1.45409  & 0.79249 
&-1.96723&1.35743\\ 
0.02   &-1.90345  &1.13828  &-1.43337  & 0.78995 
&-1.92173&1.35336\\ 
0.04   &-1.87614  &1.13198  &-1.41278  & 0.78730 
&-1.87639&1.35319\\ 
0.06   &-1.84906  &1.12535  &-1.39234  & 0.78455 
&-1.83121&1.35092\\ 
0.08   &-1.82221  &1.11838  &-1.37206  & 0.78167 
&-1.78621&1.34853\\ 
0.10   &-1.79561  &1.11099  &-1.35194  & 0.77866 
&-1.74140&1.34604\\ 
0.12   &-1.76928  &1.10319  &-1.33199  & 0.77551 
&-1.69679&1.34339\\ 
0.14   &-1.74325  &1.09489  &-1.31223  & 0.77272 
&-1.65239&1.34061\\ 
0.16   &-1.71755  &1.08608  &-1.29267  & 0.76871
&-1.60822&1.33767\\
0.18   &-1.69219  &1.07668  &-1.27332  & 0.76504 
&-1.58472&\\ 
0.20   &-1.66722  &1.06662  &-1.25420  & 0.76115     & 
&\\ 
0.22   &-1.64267  &1.05582  &-1.23533  & 0.75702     & 
&\\ 
0.24   &-1.61859  &1.04418  &-1.220       &             & 
&\\ 
0.26   &-1.59502  &1.03157  &-1.214       &             & 
&\\ 
0.28   &-1.5720   &1.01785  &-1.212       &             & 
&\\ 
\end{tabular} 
 \end{table} 
\begin{table} 
\caption{ DMRG results for the transverse spin-spin 
correlation length $\xi$ in the systems $(S_1,S_2)=(3/2,1)$, 
$(1,1/2)$, and $(3/2,1/2)$ 
for different values of the 
frustration parameter $J$.} 
\label{table2} 
\begin{tabular}{cccccccccccc} 
J &0.00  &0.02 &0.04 &0.06 &0.08 &0.10 &0.12 &0.14 &0.16 &0.18 &0.20\\
\tableline 
(3/2,1) 
&1.74&1.67&1.60&1.53&1.45&1.37&1.29&1.20&1.11&1.01&0.89\\
(1,1/2) 
&1.01&0.97&0.93&0.89&0.84&0.80&0.74&0.69&0.61&0.46&0.50\\
(3/2,1/2) 
&0.75&0.72&0.67&0.64&0.56&0.53&0.45&     &     &     &
\end{tabular} 
 \end{table} 
\begin{table} 
\caption{ DMRG results for the disorder point $J_D$ and 
the first-order transition 
point  $J_T$ found by DMRG in the systems $(S_1,S_2)=(3/2,1)$, 
$(1,1/2)$, and $(3/2,1/2)$. 
 $J_C=w/2(w^2+1)$ is  classical commensurate-incommensurate 
 transition point. $J_D^{sw}=1/2(w+1)$ is the LSWT result for the
 disorder point.} 
\label{table3} 
\begin{tabular}{ccccc} 
$(S_1,S_2)$&$J_D^{sw}$&$J_D(DMRG)$ 
&$J_C$&$J_T(DMRG)$\\ 
\tableline 
(3/2,1)&1/5&0.235(5)
&3/13&0.280(1)\\
(1,1/2)&1/6&0.177(2)
&1/5&0.231(1)\\
(3/2,1/2)&1/8&0.14(2)
&3/20&0.161(1)
\end{tabular} 
 \end{table} 
\end{document}